# Suppression of antiferromagnetic order in low-dopped ceramic samples La$_{2-x}$Sr$_x$CuO$_4$.

N.V. Dalakova[1], B.I. Belevtsev[1], E.Yu. Beliayev[1], A.S. Panfilov[1], N.P. Bobrysheva[2]

[1]*B.Verkin Institute for Low Temperature Physics and Engineering, NAS of Ukraine*
[2]*St. Petersburg State University, Department of Chemistry, St. Petersburg 198504, Russia*

**Possible influence of spatially inhomogeneous distribution of strontium impurities on the temperature of antiferromagnetic (AFM) ordering in ceramic samples La$_{2-x}$Sr$_x$CuO$_4$ has been revealed.**

E-mail: dalakova@ilt.kharkov.ua

The parent compound for La$_{2-x}$Sr$_x$CuO$_4$ is a stoichiometric lanthanum cuprate La$_2$CuO$_4$, which is a hard Mott-Hubbard AFM insulator with the Neel temperature $T_N \approx (300 \div 320 \text{ K})$ [1]. Alloying La$_2$CuO$_4$ with excess oxygen or partial replacing the lanthanum atoms by divalent atoms of alkaline earth metals (Ca, Sr or Ba) for producing charge carriers (holes) leads to suppression of the AFM order (decrease in $T_N$). When $x < 0.05$ La$_{2-x}$Sr$_x$CuO$_4$ remains an insulator. In the low-temperature limit ($T \to 0$) the AFM ordering persists to the strontium concentration $x = 0.02$. [2]. The concentrations $0.02 < x < 0.05$ are a region of strong antiferromagnetic correlations. It is agreed that charge carriers (holes) moving through the lattice affect the spin configuration and destroy the long-range antiferromagnetic order. As a result, the Neel temperature $T_N$ decreases. According to the known phase diagram [2] $T_N$ decreases monotonically as the concentration of impurities (charges) increases. The dependence $T_N(x)$ in [2] is based on the Hall coefficient data measured on a large number of high quality single crystals [3]. The homogeneity of the crystals was monitored carefully. Currently, ceramic materials based on superconducting cuprates are widely used in many areas (microelectronics, energy transmission and energy storage). Ceramic samples are frequently used to study the properties of cuprates in the AFM state. These materials are inherently heterogeneous structures. Along with the intrinsic phase inhomogeneity stemming from phase separation (PS) property of cuprates [4], these materials have heterogeneity associated with the structural and phase disordering at the grain boundaries. It is therefore interesting to investigate the influence of the degree of structural disorder on the temperature of AFM ordering in the series of ceramic cuprate samples.

The goal of this study was to synthesize and investigate ceramic samples of $La_{2-x}Sr_xCuO_4$ with the Sr concentration $0 < x < 0.01$. The samples were prepared by standard solid phase synthesis. It can be assumed that because of a very low Sr concentration these samples show additional unrecoverable heterogeneity associated with an external (extrinsic) factor. This type of heterogeneity depends on the preparation technology and is poorly controlled in systems with low levels of doping [5]. Macroscopically homogeneous samples may be inhomogeneous at the microscopic level. Conventional control techniques (X-ray diffraction, electron microscopy) are not accurate enough to register heterogeneity in such systems. However, the resistive, magnetic and magnetoresistive properties of these objects are very sensitive to any structural fluctuations, including the impurity distribution. These properties may be critically dependent on other factors such as the temperature and the time of synthesis, the annealing temperature or the rate of temperature rise. To obtain more reliable results two sets of $La_{2-x}Sr_xCuO_4$ samples have been prepared and investigated. The samples of each set were prepared in a single technological cycle under the same conditions. This allowed us to avoid uncontrolled variations of the properties of the samples within one set that might be caused by minor differences in the synthesis conditions. At the final stage of synthesis our tablets were calcinated at $T \approx 960°$ C. The calcination time was 50 hours for the first set of samples, and 67 hours for the second one. At the stage of preparation all the samples were tested by X-ray, magnetic and electron microscopic methods. The microstructure of the samples and the elemental composition of the individual composition phases were determined by scanning electron microscopy using a SEM Cam Scan microscope. The copper and lanthanum contents were investigated with an EDS LINK AN-10000 spectrometer. The strontium content was determined in five areas of the sample using a high-resolution WDS MIKROSPEC spectrometer. The average error in the element contents was ± 0.3 wt. % and varied depending on the atomic number of the element. The grain size of the ceramic was $\approx$ 10 microns. The cross-sectional micrographs taken on two samples of $La_{2-x}Sr_xCuO_4$ of the second set with $x = 0$ and $x = 0.002$ are shown in Fig. 1. The analysis of the second set of the samples shows some depositions of the CuO phase (dark spots on the micrographs). These were absent in the samples of the first set. Thus, we conclude that the samples of the first set are more homogeneous.I

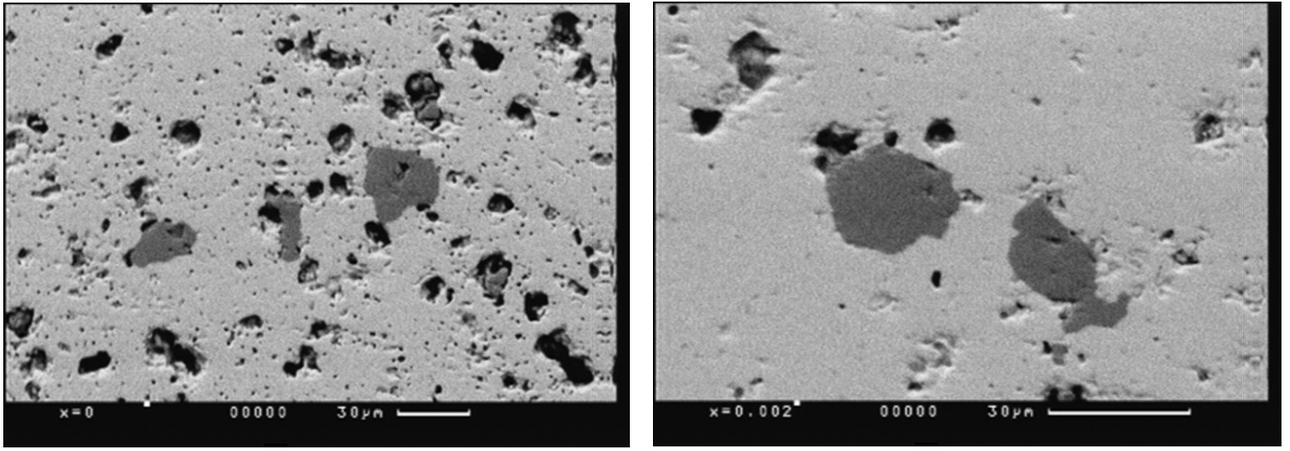

Fig.1. Micrographs of thin sections of samples La$_{2-x}$Sr$_x$CuO$_4$ with $x = 0$ (left) and $x = 0.002$ (right)

The results of X-ray micrographic spectral analysis are presented in Table 1. The Sr contents correspond to the nominal value $x = 0.01$ in the samples La$_{1.99}$Sr$_{0.01}$CuO$_4$ and are inconsistent with the prescribed concentration in the test area of the samples La$_{1.999}$Sr$_{0.001}$CuO$_4$, La$_{1.998}$Sr$_{0.002}$CuO$_4$ and La$_{1.995}$Sr$_{0.005}$CuO$_4$. This indicates a higher degree of composition heterogeneity of the latter samples. The degree of compositional (and charge) inhomogeneity increases with a decreasing Sr concentration.

The temperature dependence of the magnetic susceptibility was measured in a magnetic field of 0.83 Tesla using a Faraday magnetometer. The dependence of the magnetic susceptibility on Sr concentration taken on the samples of both sets is illustrated in Fig. 2. The dotted lines in Figs. 2a and 2b correspond to the known line of the AFM transition in the phase diagram of La$_{2-x}$Sr$_x$CuO$_4$ [2]. The Neel temperature $T_N$ of the samples of the first set with the Sr contents $x = 0.005$ and $x = 0.01$ corresponds to the phase diagram. In two samples La$_{1.999}$Sr$_{0.001}$CuO$_4$ the AFM transition occurs at much lower temperature (~212 K) than that following from the phase diagram (Fig. 2a). To determine the degree of magnetic homogeneity, magnetic susceptibility of the samples of the first set was measured on five samples for each concentration. $T_N$ varied slightly at all concentrations. The variation was ± 7% at $x = 0.001$ and $x = 0.01$ and ± 4 % at $x = 0.005$. At the same time, $T_N$ decreased significantly in comparison with the AFM transition line on the phase diagram in the samples with the lowest Sr content x = 0.001. This deviation is approximately 90 K, and can hardly be attributed to the magnetic inhomogeneity recorded in magnetic measurements. The $T_N$ values measured on the samples of set two (Fig. 2b) do not coincide with the phase diagram line [2]. The lower $T_N$ in samples of the second set may be associated with precipitation of CuO phase ( Fig.1). These precipitation enrich the matrix by strontium and oxygen, and consequently increase the effective concentration of the charge carriers. As a result, the temperature $T_N$ decreases (dashed line in Fig. 2b). This deviation from the monotonic behaior of $T_N(x)$ is also observed in the samples of the second set at extremely low concentrations ($x = 0.001$, $x = 0.002$).

Table 1

The average contents of CuO, $La_2O_3$, SrO and the standard deviations from the average Sr content (ASC) for two sets of of $La_{2-x}Sr_xCuO_4$ samples.

| sample | CuO | $La_2O_3$ | SrO | ASC Sr | Sum |
|---|---|---|---|---|---|
| The first set of samples | | | | | |
| $La_{1.999}Sr_{0.001}CuO_4$ | 1.03 | 1.97 | 0.0020 | 0.0003 | 3.00 |
| $La_{1.995}Sr_{0.005}CuO_4$ | 1.01 | 2.00 | 0.0019 | 0.0002 | 3.00 |
| $La_{1.99}Sr_{0.01}CuO_4$ | 0.97 | 2.02 | 0.01 | 0.0000 | 3.00 |
| The second set of samples | | | | | |
| $La_2CuO_4$ | 1.00 | 2.00 | 0.00 | — | 3.00 |
| $La_{1.999}Sr_{0.001}CuO_4$ | 0.98 | 2.01 | 0.0122 | 0.0015 | 3.00 |
| $La_{1.998}Sr_{0.002}CuO_4$ | 1.00 | 2.00 | 0.00 | — | 3.00 |
| $La_{1.995}Sr_{0.005}CuO_4$ | 1.10 | 1.90 | 0.0081 | 0.0012 | 3.00 |
| $La_{1.99}Sr_{0.01}CuO_4$ | 0.99 | 2.00 | 0.01 | 0.0000 | 3.00 |

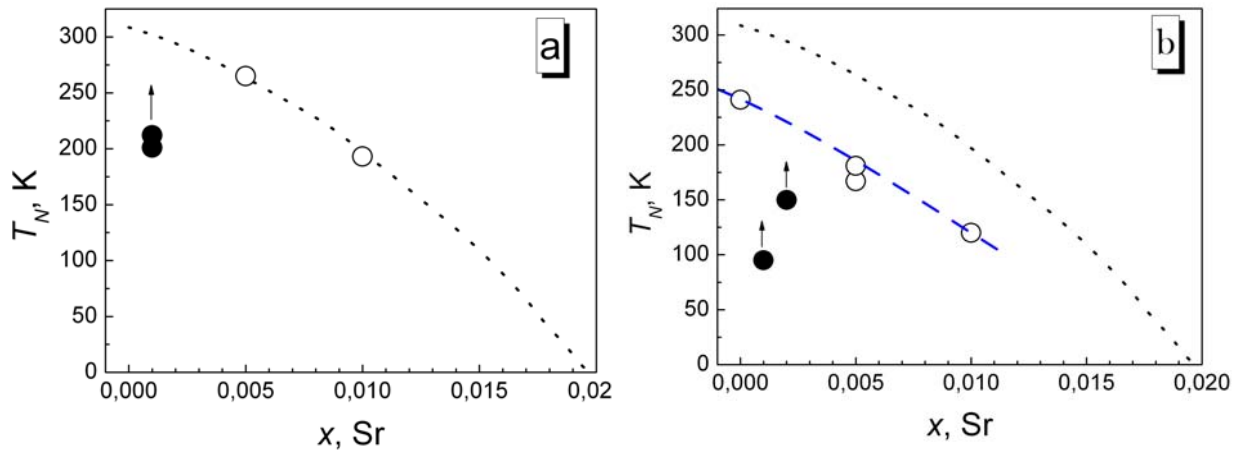

Fig.2. The dependence of $T_N$ on the Sr concentration: a - the first set of samples; b - the second set of samples.

The temperature dependence of the resistance was measured at the current $J = 1$ μA for the samples of the first set and $J = 1000$ μA for the samples of the second set. The results are shown in Fig.3. The samples of the first set are more dielectric. Their resistivity ρ is several orders of magnitude higher than ρ in the samples of second set. This result corresponds to measured $T_N(x)$ and the data in Table 1. Samples of the first and the second sets exhibit different temperature dependences $\rho(T)$. In the interval $T \approx 20 \div 100$ K the sample from the first set obey the Mott's law of variable range hopping conduction. But the samples of the second set do not follow this law, although at $T < 50$ K their resistance increases exponentially. The qualitative difference in the dependence $\rho(T)$ between the samples of the first set and the second set is due to the slight distinctions in the conditions of synthesis. We have $\rho(x = 0.001) < \rho(x = 0.005)$ for the samples of both sets. It is essential that the localization lengths for the samples of the first set with $x = 0.001$ and $x = 0.005$ are close: $L_c \approx 0.3$ nm ($x = 0.001$), $L_c \approx 0.28$ nm ($x = 0.005$). Thus, the results of electron microscopy, the electron microprobe analysis (Table 1) and the measured resistivities demonstrate good correlation. At the same time, the results of magnetic measurements of $T_N(x)$ show an anomaly at the concentrations $0 < x < 0.005$: $T_N$ decreases relative to the phase diagram line. Besides, the deviation of $T_N$ from the phase diagram line tends to increase as the nominal concentration $x$ decreases and the degree of structural disorder associated with compositional inheterogeneity of the samples increases. This behavior is unusual. And until now there has been no unambiguous explanation for the observed anomaly. Nevertheless, it is clear that the anomaly is related to the compositional heterogeneity of the samples.

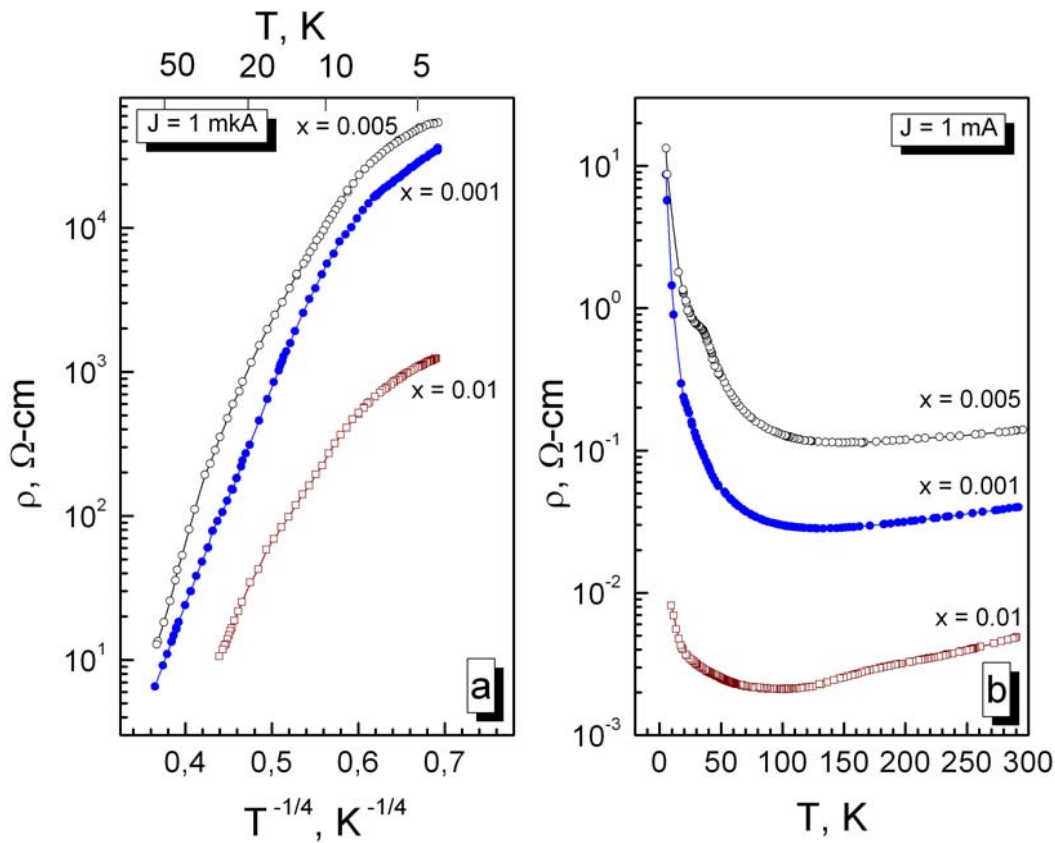

Figure 3. The temperature dependence of resistivity: a – the first set of samples
b – the second set of samples.

Ceramic cuprate samples are granular systems with non-universal structural disorder such as clusters of impurities and precipitations of another phase. It is noted [6] that even a slight structural disorder can affect the nature of the phase transition. When the structural disorder is stronger than thermal or quantum fluctuations, it causes a significant smearing of the phase transition. These ideas were developed in [6] to describe the superconducting phase transition temperature in a superconductor with a structural disorder. We assume that structural fluctuations can influence in a similar way the magnetic phase transition. A sufficiently strong increase in the degree of structural disorder can cause smearing of the phase transition and reducing the Neel temperature.

In the limiting case $x = 0$ ($T_N$ = 320 K), when each site is occupied by a single particle, the system of strongly correlated electrons is strongly degenerate in spin variable. According to [7], removing even a single particle (adding one hole) we translate such a system in the ferromagnetic state. In this case $T_N \to 0$. It is possible that in the system of AFM granules an addition of a minor amount of Sr suppresses the AFM order in individual granules, and the temperature $T_N$ of the entire system decreases. The interaction between the spin system and the charge carriers may be an additional factor of suppression of AFM. It is known that the motion of an isolated hole in the antiferromagnet is frustrated [7]. The motion of an isolated hole in antiferromagnetic media causes dynamic spin fluctuations and forms a chains of ferromagnetic bonds. This should also lead to a decrease in the Neel temperature.